\documentclass{PoS}

\vbox{
\hbox{ADP-05-18/T628}
}

\PoS{PoS(LAT2005)040}

\ShortTitle{Light-Quark FLIC Fermion Simulations of the $1^{-+}$ Exotic Meson }
\title{Light-Quark FLIC Fermion Simulations of the $1^{-+}$ Exotic Meson }

\author{J.~N.~Hedditch\footnote{corresponding author}, B.~G.~Lasscock, D.~B.~Leinweber, \speaker{A.~G.~Williams}\\

Department of Physics and Mathematical Physics and\\
        Special Research Centre for the
        Subatomic Structure of Matter,                          \\
        University of Adelaide, 5005, Australia\\
        E-mail: \email{jhedditc@physics.adelaide.edu.au}}

\author{W.~Kamleh\\

School of Mathematics,
Trinity College\\
Dublin 2,
Ireland}

\author{J.~M.~Zanotti\\
John von
    Neumann-Institut f\"ur Computing NIC, \\
    Deutches Elektronen-Synchrotron DESY, \\ 
    D-15738 Zeuthen, Germany
}

\abstract{We investigate the mass of the $1^{-+}$ exotic meson, created with hybrid
interpolating fields. Access to light quark masses approaching 25
MeV is facilitated by the use of the Fat-Link Irrelevant Clover (FLIC)
fermion action, and large ($20^3 \times 40$) lattices.
Our results indicate that the $1^{-+}$ exotic exhibits
significant curvature close to the chiral limit,
and yield a $1^{-+}$ mass in agreement with the $\pi_1 (1600)$ candidate and
exclusive of the $\pi_1 (1400)$.}

\FullConference{XXIIIrd International Symposium on Lattice Field Theory\\

		 25-30 July 2005\\

		 Trinity College, Dublin, Ireland}

\begin{document}

%%%%%%%%%%%%%%%%%%%%%%%%%%%%%%%%%%%%%%%%%%%%%%%%%%%%%%%%%%%%%%%%%%%%%%%%%%
\section{Introduction}

The exotic mesons comprise a rare
vehicle for the elucidation of the relatively unexplored role of
gluons in QCD.  The Particle Data Group \cite{Eidelman:2004wy} reports two
candidates for the $1^{-+}$ exotic, the $\pi_1 (1400)$ at $1.376(17)
{\rm GeV}$, and the $\pi_1 (1600)$ at $1.596^{+25}_{-14}\ {\rm GeV}$.
The interpretation of the experimental data continues to inspire
discussion~\cite{Lu:2004yn,Dzierba:2005sr}.

Michael~\cite{Michael:2003xg} provides a good summary of lattice results
in this field up to 2003,
concluding that the light-quark exotic is predicted by lattice studies
to have a mass of $1.9(2)$ GeV, which differs from both experimental
candidates. It should be emphasised, however, that previous results are
derived from extrapolations from relatively heavy quark masses.

In order to minimize the need for extrapolation one requires access to
quark masses near the chiral regime on large physical volumes.  Our
study considers a physical volume of $(2.6\ \rm fm)^3$, and the ${\mathcal
O}(a)$-improved FLIC fermion action
\cite{Zanotti:2002ax,Zanotti:2004dr} whose improved chiral properties
\cite{Boinepalli:2004fz} permit the use of very light quark masses
which are key to our results.

%%%%%%%%%%%%%%%%%%%%%%%%%%%%%%%%%%%%%%%%%%%%%%%%%%%%%%%%%%%%%%%%%%%%%%%%%%
\section{Lattice Simulations}

%\subsection{Interpolating Fields}

We use local interpolating fields, coupling colour-octet quark
bilinears to chromo-electric and chromo-magnetic fields. 
It is possible to generalise the
interpolating fields to include non-local components where link paths
are incorporated to maintain gauge invariance and carry the nontrivial
quantum numbers of the gluon fields \cite{Lacock:1996vy,Lacock:1998be}.
Such an approach does not lead to an increase in signal
for the ground state $1^{-+}$ exotic commensurate with the increased
computational cost of multiple fermion-matrix inversions.  

Gauge-invariant Gaussian smearing
\cite{Gusken:1989qx,Zanotti:2003fx} is applied at the fermion source
($t=8$), and local sinks are used to maintain strong signal in the
two-point correlation functions.  
In this work we considered four interpolating fields for the $1^{-+}$
exotic:
\begin{equation}
\chi_1 = \bar{q}^a \gamma_4 E^{ab}_j q^{b},
\end{equation}
\begin{equation}
\chi_2 = i \epsilon_{jkl} \bar{q}^a \gamma_k B^{ab}_l q^b,
\end{equation}
\begin{equation}
\chi_3 = i \epsilon_{jkl} \bar{q}^a \gamma_4 \gamma_k B^{ab}_l q^b,
\end{equation}
and 
\begin{equation}
\chi_4 = \epsilon_{jkl} \bar{q}^a \gamma_5 \gamma_4 \gamma_k E^{ab}_l q^b \, .
\end{equation}

The interpolating fields which couple large-large and small-small
spinor components (i.e $\chi_2$ and $\chi_3$) provide the strongest signal for the
$1^{-+}$ state.  

%\subsection{Lattice Field Strength Tensor}

In order to obtain the chromo-electric and chromo-magnetic fields with
which we build the hybrid operators, we make use of a modified version
of APE smearing~\cite{ape}, in which the smeared links do not involve averages
which include links in the temporal direction.  In this way we
preserve the notion of a Euclidean `time' and avoid overlap of the creation and
annihilation operators.  
In this study, the smearing fraction $\alpha = 0.7$ (keeping 0.3 of
the original link) and the process of smearing and $SU(3)$ link
projection is iterated four times \cite{Bonnet:2000dc}.
Smearing the links permits the use of highly improved definitions of
the lattice field strength tensor, from which our hybrid operators are
derived.  Details of the ${\mathcal O(a^4)}$-improved tensor are given
in~\cite{Bilson-Thompson:2002jk}.  

%%%%%%%%%%%%%%%%%%%%%%%%%%%%%%%%%%%%%%%%%%%%%%%%%%%%%%%%%%%%%%%%%%%%%%%%%%

%\subsection{Fat-Link Irrelevant Fermion Action}

Propagators are generated using the fat-link irrelevant clover (FLIC)
fermion action \cite{Zanotti:2004dr} where the irrelevant Wilson and
clover terms of the fermion action are constructed using APE-Smeared links~\cite{ape},
while the relevant operators use the untouched (thin) gauge links.
In the FLIC action, this yields improved chiral properties and
reduces the problem of exceptional configurations
encountered with clover actions \cite{Boinepalli:2004fz}, and
minimizes the effect of renormalization on the action improvement
terms \cite{Leinweber:2002bw}.  
Details of this approach may be found in
reference \cite{Zanotti:2004dr}.  FLIC fermions provide a new form of
nonperturbative ${\mathcal O}(a)$ improvement
\cite{Boinepalli:2004fz,Leinweber:2002bw} where near-continuum results
are obtained at finite lattice spacing.

%%%%%%%%%%%%%%%%%%%%%%%%%%%%%%%%%%%%%%%%%%%%%%%%%%%%%%%%%%%%%%%%%%%%%%%%%%
%

\begin{figure}[tb]
\begin{center}
$\begin{array}{c@{\hspace{1in}}c}
\multicolumn{1}{l}{\hspace{-5mm}\mbox{}} &
	\multicolumn{1}{l}{\mbox{}} \\ [-0.53cm]
\hspace{-2mm}\includegraphics[height=0.45\hsize,angle=90]{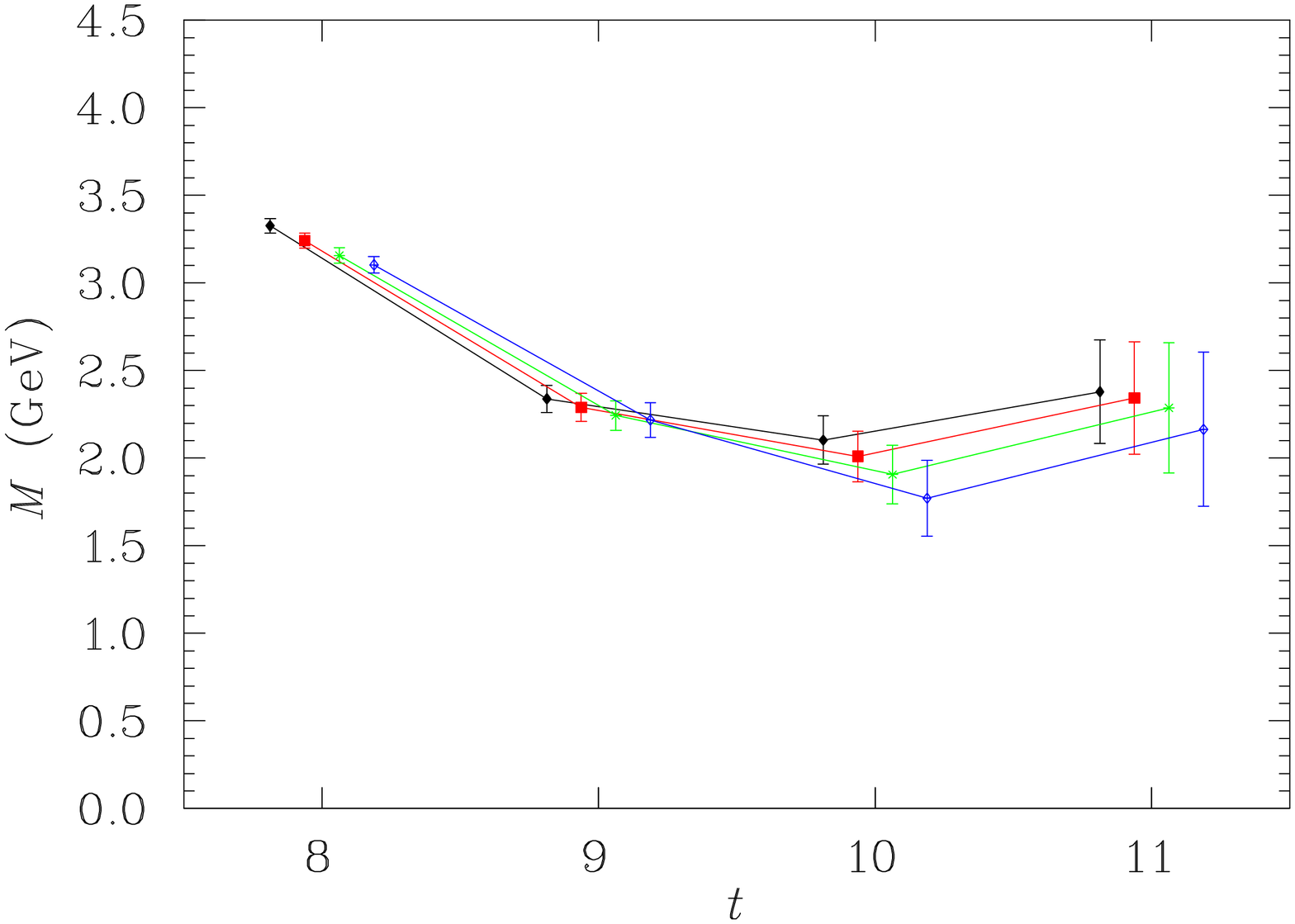} &
\hspace{-20mm} \includegraphics[height=0.45\hsize,angle=90]{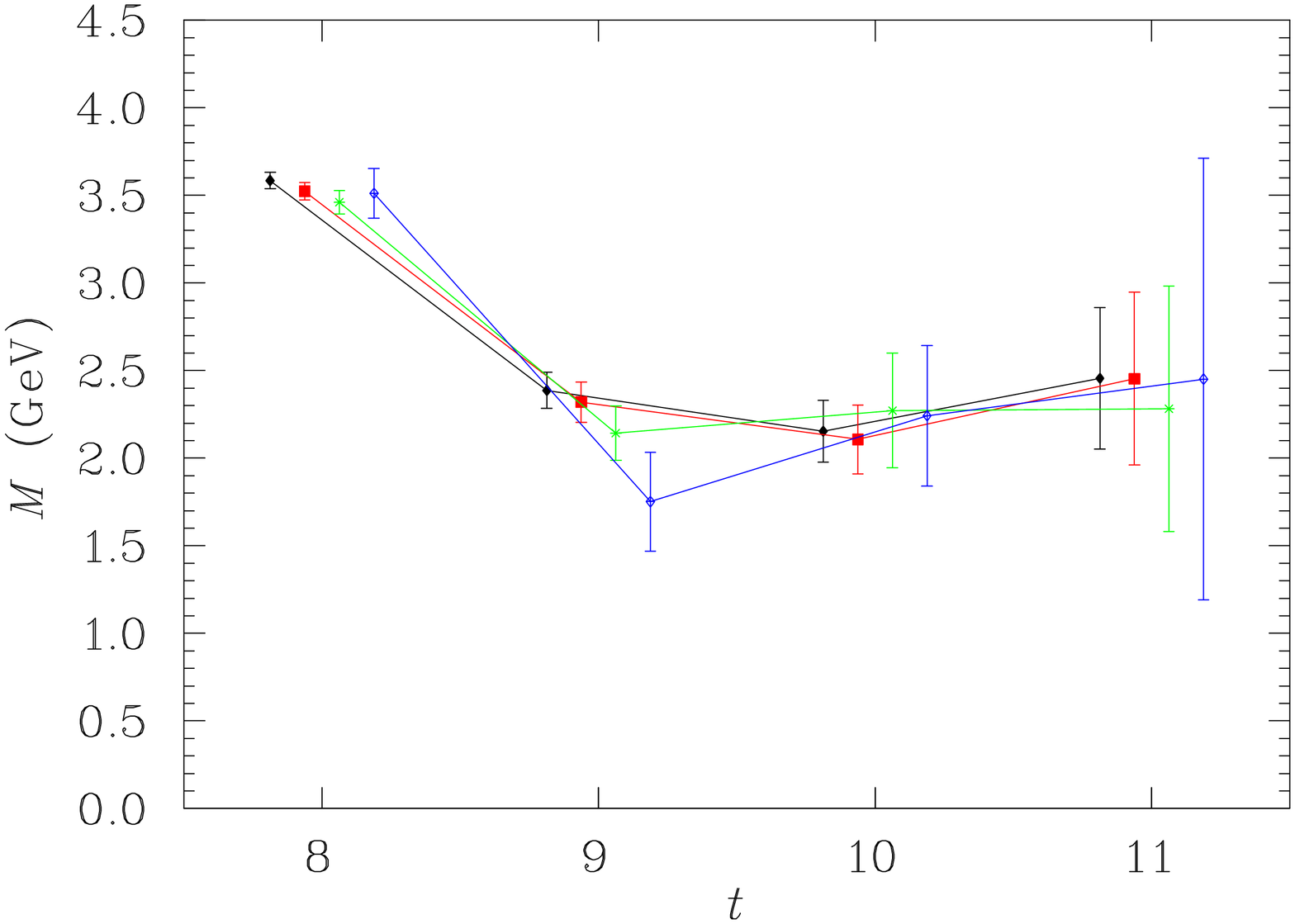} \\ [0.4cm]
\mbox{\bf (a)} & \mbox{\bf (b)} 
\end{array}$
\end{center}
\caption{ Effective masses extracted with interpolators $\chi_2$ (a) and $\chi_3$ (b).}
\label{onemptwothree}
\end{figure}

\begin{figure}[hbt]
\includegraphics[height=0.9\hsize,angle=90]{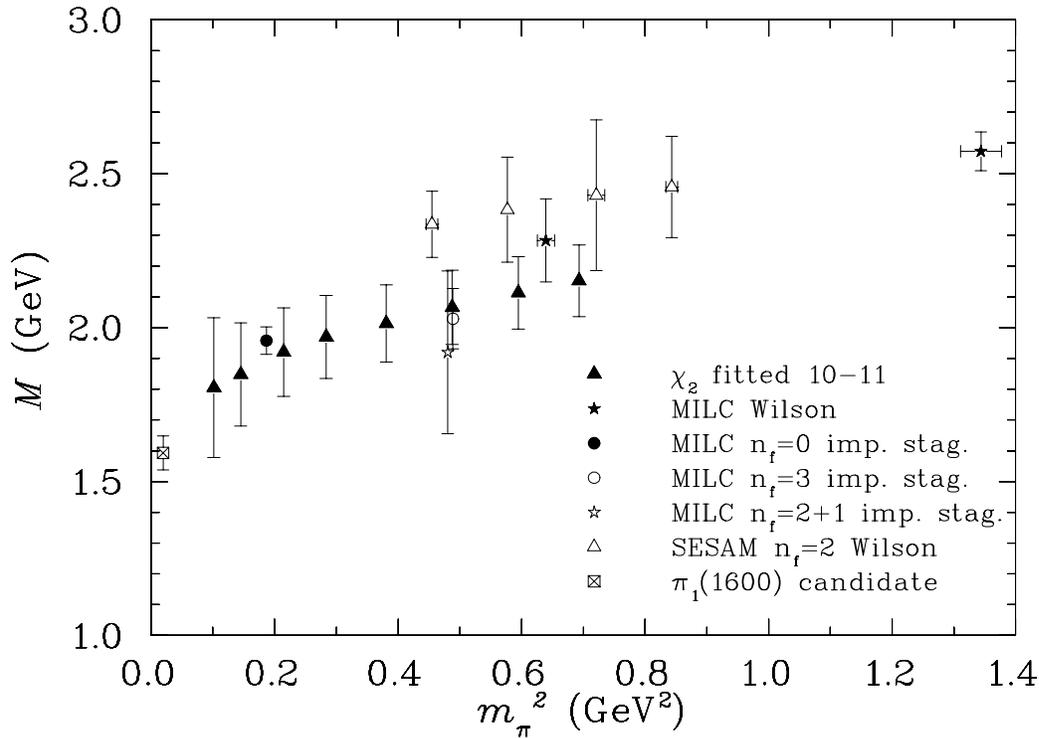}
\caption{\label{survey} A survey of results in this field. Open and closed symbols denote dynamical and quenched simulations respectively. The MILC results are taken from~\cite{Bernard:1997ib} and
show their $Q^4,1^{-+} \to 1^{-+}$ results, fitted from $t=3$ to $t=11$. }
\end{figure}

%\subsection{Gauge Action}

We use quenched-QCD gauge fields created by the CSSM Lattice
Collaboration with the ${\mathcal O}(a^2)$ mean-field improved
L\"uscher-Weisz plaquette plus rectangle gauge action
\cite{Luscher:1984xn} using the plaquette measure for the mean link.
The CSSM configurations are generated using the Cabibbo-Marinari
pseudo-heat-bath algorithm~\cite{Cabibbo:1982zn} using a parallel algorithm
with appropriate link partitioning \cite{Bonnet:2000db}.  To improve
the ergodicity of the Markov chain process, the three diagonal SU(2)
subgroups of SU(3) are looped over twice~\cite{Bonnet:2001rc} and a
parity transformation \cite{Leinweber:2003sj} is applied randomly to
each gauge field configuration saved during the Markov chain process.

%\subsection{Simulation Parameters}

The calculations of meson masses are performed on $20^3\times 40$
lattices at $\beta=4.53$, which provides a lattice spacing of $a =
0.128(2)$~fm set by the Sommer parameter $r_0=0.49\ \rm fm$.
A fixed boundary condition in the time direction is used for the fermions
by setting $U_t(\vec x, N_t) = 0\ \forall\ \vec x$ in the hopping terms
of the fermion action, with periodic boundary conditions imposed in the
spatial directions.
Eight quark masses are considered in the calculations and the strange
quark mass is taken to be the third heaviest quark mass.  This
provides a pseudoscalar mass of 697 MeV which compares well with the
experimental value of $( 2M_K^2 - M_\pi^2 )^{1/2} = 693\, {\rm MeV}$
motivated by leading order chiral perturbation theory.  
The analysis is based on a sample of 345 configurations, and the error
analysis is performed by a third-order single-elimination jackknife,
with the $\chi^2$ per degree of freedom (${\chi^2}/dof$) obtained via
covariance matrix fits.

\section{Results}

%%%%%%%%%%%%%%%%%%%%%%%%%%%%%%%%%%%%%%%%%%%%%%%%%%%%%%%%%%%%%%%%%%%%%%%%%%
%
\begin{table}[t]
\begin{center}
\caption{\label{masstable}$1^{-+}$ Exotic Meson mass (GeV) vs square
  of pion mass (GeV$^2$).} 
\begin{tabular}{c|ll|ll|ll}
\hline
 & \multicolumn{2}{c}{$\chi_2$ fit 10-11} & \multicolumn{2}{c}{$\chi_2$ fit 10-12} & \multicolumn{2}{c}{$\chi_3$ fit 10-11} \\
\hline
 $m_{\pi}^2$& $m$ & $\chi^2/dof \ $ & $m$ & $\chi^2/dof \ $ & $m$ & $\chi^2/dof$ \  \\
\hline
0.693(3) & 2.15(12) & 0.69 & 2.16(11) & 0.44 & 2.20(15) & 0.45 \\
0.595(4) & 2.11(12) & 0.77 & 2.12(11) & 0.51 & 2.18(16) & 0.46\\
0.488(3) & 2.07(12) & 0.85 & 2.08(12) & 0.59 & 2.15(17) & 0.41\\
0.381(3) & 2.01(12) & 0.91 & 2.03(12) & 0.65 & 2.14(19) & 0.29\\
0.284(3) & 1.97(13) & 0.78 & 1.98(13) & 0.55 & 2.27(29) & 0.0001\\
0.215(3) & 1.92(14) & 0.78 & 1.92(14) & 0.40 & 2.25(31) & 0.02\\
0.145(3) & 1.85(17) & 0.57 & 1.84(17) & 1.76 & 2.26(37) & 0.02\\
0.102(4) & 1.80(23) & 0.13 & 1.75(23) & 3.04 & 2.46(58) & 0.03\\
\hline
\end{tabular}
\end{center}
\end{table}

Figure~\ref{onemptwothree} shows the effective mass for
the two preferred interpolators.  For clarity, we have plotted the
results for every second quark mass used in our simulation.  
The effective masses exhibit plateaus at $0.256$~fm from the source
which is 
consistent with Ref.~\cite{Bernard:1997ib}, where a similar effect is
seen after approximately  $0.21$ to $0.28$~fm.

Table~\ref{masstable} summarizes our results for the mass of the
$1^{-+}$ meson, with the squared pion-mass provided as a measure of
the input quark mass.  
The agreement between the interpolators is significant, as we expect 
them to posess considerably different excited-state
contributions, based on experience with pseudoscalar interpolators
\cite{Holl:2005vu}.

Fig.~\ref{survey} summarizes a collection of results for
the mass of the $1^{-+}$ obtained in lattice QCD simulations thus far.
The current results presented herein (full triangles) are compared with results from
the MILC \cite{Bernard:1997ib,Bernard:2002rz} and SESAM \cite{Lacock:1998be}
collaborations, both of which provide a consistent scale via $r_0$.

We perform a linear fit to the $1^{-+}$ mass using the four lightest quark masses and
a quadratic form to all 8 masses. Systematic uncertainties associated with
chiral nonanalytic curvature are estimated at 50 MeV\cite{InPrep,Thomas:2001gu}. A third-order
single-elimination jackknife error analysis yields masses of $1.74(24)$
and $1.74(25)$ GeV for the linear and quadratic fits, respectively.
These results agree within one standard deviation with the
experimental $\pi_1 (1600)$ result of $1.596^{+25}_{-14}\ {\rm GeV}$, and exclude the
mass of the $\pi_1 (1400)$ candidate.

%%%%%%%%%%%%%%%%%%%%%%%%%%%%%%%%%%%%%%%%%%%%%%%%%%%%%%%%%%%%%%%%%%%%%%%%%%
\section{Conclusion}

We have found a compelling signal for the $J^{PC}=1^{-+}$ exotic meson
at very light quark masses, from which we can extrapolate a physical
mass of $1.74(24)$ GeV.  Thus for the first time in lattice studies,
we find a $1^{-+}$ mass in agreement with the $\pi_1 (1600)$
candidate.

Looking forward, it will be important to quantify the effects of the
quenched approximation.  Of particular interest will be
the extent to which the curvature observed in approaching the chiral
regime is preserved in full QCD.

Whilst the rapidity with which
we establish a plateau in our effective mass plots suggests that our
current fermion operator smearing is near optimal for isolating the
ground state, it might be possible to reduce the statistical errors
through a careful selection of parameters coming out of a systematic
exploration of the parameter space.

%%%%%%%%%%%%%%%%%%%%%%%%%%%%%%%%%%%%%%%%%%%%%%%%%%%%%%%%%%%%%%%%%%%%%%%%%%
\begin{acknowledgments}

We thank Doug Toussaint for sharing his collection of results for the
$1^{-+}$ mass.  We thank the Australian Partnership for Advanced Computing
(APAC) the South Australian Partnership for Advanced
Computing (SAPAC) for generous grants of supercomputer time which have
enabled this project.  This work was supported by the Australian
Research Council.

\end{acknowledgments}


\begin{thebibliography}{40}

%\cite{Eidelman:2004wy}
\bibitem{Eidelman:2004wy}
  S.~Eidelman {\it et al.}  [Particle Data Group],
  %``Review of particle physics,''
  Phys.\ Lett.\ B {\bf 592}, 1 (2004).
  %%CITATION = PHLTA,B592,1;%%

%\cite{Lu:2004yn}
\bibitem{Lu:2004yn}
  M.~Lu {\it et al.}  [E852 Collaboration],
  %``Exotic meson decay to omega pi0 pi-,''
  Phys.\ Rev.\ Lett.\  {\bf 94}, 032002 (2005)
  [arXiv:hep-ex/0405044].
  %%CITATION = HEP-EX 0405044;%%

%\cite{Dzierba:2005sr}
\bibitem{Dzierba:2005sr}
  A.~R.~Dzierba, R.~Mitchell, A.~P.~Szczepaniak, M.~Swat and S.~Teige,
  %``A search for J(PC) = 1-+ exotic mesons in the pi- pi- pi+ and pi- pi0  pi0
  %systems,''
  arXiv:hep-ex/0502022.
  %%CITATION = HEP-EX 0502022;%%

%\cite{Zanotti:2002ax}
\bibitem{Zanotti:2002ax}
J.~M.~Zanotti {\it et al.},
%``Novel fat-link fermion actions,''
Nucl.\ Phys.\ Proc.\ Suppl.\  {\bf 109A}, 101 (2002)
[arXiv:hep-lat/0201004].
%%CITATION = HEP-LAT 0201004;%%

\bibitem{Zanotti:2004dr}
  J.~M.~Zanotti, B.~Lasscock, D.~B.~Leinweber and A.~G.~Williams,
  %``Scaling of FLIC fermions,''
  Phys.\ Rev.\ D {\bf 71}, 034510 (2005)
  [arXiv:hep-lat/0405015].
  %%CITATION = HEP-LAT 0405015;%%

%\cite{Boinepalli:2004fz}
\bibitem{Boinepalli:2004fz}
  S.~Boinepalli, W.~Kamleh, D.~B.~Leinweber, A.~G.~Williams and
  J.~M.~Zanotti,
  %``Improved chiral properties of FLIC fermions,''
  Phys.\ Lett.\ B {\bf 616}, 196 (2005)
  [arXiv:hep-lat/0405026].
  %%CITATION = HEP-LAT 0405026;%%


%\cite{Lacock:1996vy}
\bibitem{Lacock:1996vy}
  P.~Lacock, C.~Michael, P.~Boyle and P.~Rowland  [UKQCD Collaboration],
  %``Orbitally excited and hybrid mesons from the lattice,''
  Phys.\ Rev.\ D {\bf 54}, 6997 (1996)
  [arXiv:hep-lat/9605025].
  %%CITATION = HEP-LAT 9605025;%%

%\cite{Bernard:1997ib}
\bibitem{Bernard:1997ib}
  C.~W.~Bernard {\it et al.}  [MILC Collaboration],
  %``Exotic mesons in quenched lattice QCD,''
  Phys.\ Rev.\ D {\bf 56}, 7039 (1997)
  [arXiv:hep-lat/9707008].
  %%CITATION = HEP-LAT 9707008;%%

%\cite{Lacock:1998be}
\bibitem{Lacock:1998be}
  P.~Lacock and K.~Schilling  [TXL collaboration],
  %``Hybrid and orbitally excited mesons in full QCD,''
  Nucl.\ Phys.\ Proc.\ Suppl.\  {\bf 73}, 261 (1999)
  [arXiv:hep-lat/9809022].
  %%CITATION = HEP-LAT 9809022;%%

%\cite{Bernard:2002rz}
\bibitem{Bernard:2002rz}
  C.~Bernard {\it et al.},
  %``Exotic hybrid mesons from improved Kogut-Susskind fermions,''
  Nucl.\ Phys.\ Proc.\ Suppl.\  {\bf 119}, 260 (2003)
  [arXiv:hep-lat/0209097].
  %%CITATION = HEP-LAT 0209097;%%

%\cite{Michael:2003xg}
\bibitem{Michael:2003xg}
  C.~Michael,
  %``Hybrid mesons from the lattice,''
  [arXiv:hep-ph/0308293].
  %%CITATION = HEP-PH 0308293;%%

%\cite{Zanotti:2001yb}
\bibitem{Zanotti:2001yb}
  J.~M.~Zanotti {\it et al.}  [CSSM Lattice Collaboration],
  %``Hadron masses from novel fat-link fermion actions,''
  Phys.\ Rev.\ D {\bf 65}, 074507 (2002)
  [arXiv:hep-lat/0110216].
  %%CITATION = HEP-LAT 0110216;%%

%\cite{Gusken:1989qx}
\bibitem{Gusken:1989qx}
  S.~Gusken,
  %``A Study Of Smearing Techniques For Hadron Correlation Functions,''
  Nucl.\ Phys.\ Proc.\ Suppl.\  {\bf 17}, 361 (1990).
  %%CITATION = NUPHZ,17,361;%%

%\cite{Zanotti:2003fx}
\bibitem{Zanotti:2003fx}
  J.~M.~Zanotti, D.~B.~Leinweber, A.~G.~Williams, J.~B.~Zhang,
  W.~Melnitchouk and S.~Choe 
  %``Spin-3/2 nucleon and Delta baryons in lattice QCD,''
  Phys.\ Rev.\ D {\bf 68}, 054506 (2003)
  [arXiv:hep-lat/0304001].
  %%CITATION = HEP-LAT 0304001;%%

\bibitem{ape}
  M.~Falcioni, M.~L.~Paciello, G.~Parisi and B.~Taglienti,
  %``Again On SU(3) Glueball Mass,''
  Nucl.\ Phys.\ B {\bf 251}, 624 (1985);
  %%CITATION = NUPHA,B251,624;%%
  M.~Albanese {\it et al.}  [APE Collaboration],
  %``Glueball Masses And String Tension In Lattice QCD,''
  Phys.\ Lett.\ B {\bf 192}, 163 (1987).
  %%CITATION = PHLTA,B192,163;%%

%\cite{Bilson-Thompson:2002jk}
\bibitem{Bilson-Thompson:2002jk}
  S.~O.~Bilson-Thompson, D.~B.~Leinweber and A.~G.~Williams,
  %``Highly-improved lattice field-strength tensor,''
  Annals Phys.\  {\bf 304}, 1 (2003)
  [arXiv:hep-lat/0203008].
  %%CITATION = HEP-LAT 0203008;%%

%\cite{Bonnet:2000dc}
\bibitem{Bonnet:2000dc}
  F.~D.~R.~Bonnet, P.~Fitzhenry, D.~B.~Leinweber, M.~R.~Stanford and A.~G.~Williams,
  %``Calibration of smearing and cooling algorithms in SU(3)-color gauge
  %theory,''
  Phys.\ Rev.\ D {\bf 62}, 094509 (2000)
  [arXiv:hep-lat/0001018].
  %%CITATION = HEP-LAT 0001018;%%

%\cite{Leinweber:2002bw}
\bibitem{Leinweber:2002bw}
  D.~B.~Leinweber, {\it et al.}
  % ``FLIC Fermions and hadron Phenomenology,''
  Eur.\ Phys.\ J. A {\bf 18}, 247 (2003)
  [arXiv:nucl-th/0211014]
  %%CITATION = NUCL-TH 0211014;%%

%\cite{Luscher:1984xn}
\bibitem{Luscher:1984xn}
  M.~Luscher and P.~Weisz,
  %``On-Shell Improved Lattice Gauge Theories,''
  Commun.\ Math.\ Phys.\  {\bf 97}, 59 (1985)
  [Erratum-ibid.\  {\bf 98}, 433 (1985)].
  %%CITATION = CMPHA,97,59;%%


%\cite{Cabibbo:1982zn}
\bibitem{Cabibbo:1982zn}
  N.~Cabibbo and E.~Marinari,
  %``A New Method For Updating SU(N) Matrices In Computer Simulations Of Gauge
  %Theories,''
  Phys.\ Lett.\ B {\bf 119}, 387 (1982).
  %%CITATION = PHLTA,B119,387;%%

%\cite{Bonnet:2000db}
\bibitem{Bonnet:2000db}
F.~D.~Bonnet, D.~B.~Leinweber and A.~G.~Williams,
%``General algorithm for improved lattice actions on parallel computing  architectures,''
J.\ Comput.\ Phys.\  {\bf 170}, 1 (2001)
[arXiv:hep-lat/0001017].
%%CITATION = HEP-LAT 0001017;%%


%\cite{Bonnet:2001rc}
\bibitem{Bonnet:2001rc}
  F.~D.~R.~Bonnet, D.~B.~Leinweber, A.~G.~Williams and J.~M.~Zanotti,
  %``Improved smoothing algorithms for lattice gauge theory,''
  Phys.\ Rev.\ D {\bf 65}, 114510 (2002)
  [arXiv:hep-lat/0106023].
  %%CITATION = HEP-LAT 0106023;%%

%\cite{Leinweber:2003sj}
\bibitem{Leinweber:2003sj}
  D.~B.~Leinweber, A.~G.~Williams, J.~b.~Zhang and F.~X.~Lee,
  %``Topological charge barrier in the Markov-chain of QCD,''
  Phys.\ Lett.\ B {\bf 585}, 187 (2004)
  [arXiv:hep-lat/0312035].
  %%CITATION = HEP-LAT 0312035;%%

%\cite{Holl:2005vu}
\bibitem{Holl:2005vu}
  A.~Holl, A.~Krassnigg, P.~Maris, C.~D.~Roberts and S.~V.~Wright,
  %``Electromagnetic properties of ground and excited state pseudoscalar
  %mesons,''
  [arXiv:nucl-th/0503043].
  %%CITATION = NUCL-TH 0503043;%%

%\cite{Thomas:2001gu}
\bibitem{Thomas:2001gu}
A.~W.~Thomas and A.~P.~Szczepaniak,
%``Chiral extrapolations and exotic meson spectrum,''
Phys.\ Lett.\ B {\bf 526}, 72 (2002)
[arXiv:hep-ph/0106080].
%%CITATION = HEP-PH 0106080;%%

\bibitem{InPrep}
J.~N.~Hedditch, W.~Kamleh, B.~G.~Lasscock, D.~B.~Leinweber, J.~M.~Zanotti and A.~G.~Williams,
in preparation.

\end{thebibliography}
\end{document}